\documentclass[preprint,showpacs,preprintnumbers,amsmath,amssymb]{revtex4}

\usepackage{graphicx}
\usepackage{dcolumn}
\usepackage{bm}

\begin{document}

\title{EVOLUTION OF OBSERVABLES IN A NUMERICAL KINETIC MODEL.} 

\author{N.S. Amelin}
\author{R. Lednicky}
\altaffiliation[Also at ]{Institute of Physics ASCR, Prague, 18221, Czech Republic}
\author{L. V. Malinina}
\altaffiliation[Also at ]{M.V. Lomonosov Moscow State University, D.V. Skobeltsyn
Institut of Nuclear Physics}
\author{T. A. Pocheptsov}
\affiliation{Joint Institute for Nuclear Research, Dubna, Moscow Region, 141980, Russia}
\author{Yu.M. Sinyukov}
\affiliation{Bogolyubov Institute for Theoretical Physics, Kiev, 03143, Ukraine}

\date{\today}

\begin{abstract}
The numerical solutions of the
nonrelativistic and relativistic Boltzmann equations
have been studied at various initial conditions.
Particularly, the known analytical
solution of the nonrelativistic Boltzmann equation at
spherically symmetric Gaussian initial conditions has been recovered
and the corresponding conservation
of particle spectra and correlation functions has been confirmed.
Similar conservation properties have been found also for
a more general class of solutions describing the
nonrelativistic and relativistic systems at anisotropic
initial conditions.
\end{abstract}



\maketitle

\section{\label{sec1}Introduction.}

The evolution of the matter formed in ultrarelativistic heavy ion
collisions is usually described within hydrodynamic\ and transport models.
The former allow one to incorporate the complicated evolution of the hot
and dense system (fireball) at the possible phase transitions encoded in
corresponding equation of state, while the latter make it possible to
evaluate particle spectra taking into account the nonequilibrated
character of their formation even at the early evolution stage.
However, in realistic conditions,
neither of these approaches nor their hybrids are able
to reproduce particle spectra and correlations
in A+A collisions.
At present there is no scenario of the evolution allowing one to
describe all the salient features of hadronic observables,
including the interferometry or correlation radii related to
the space-time scales of the fireball.
Particularly, the observed approximate energy
independence of the radii is considered as the interferometry puzzle.

It was noted \cite{Sinyukov}, based on an analysis of the exact solution
of the nonrelativistic Boltzmann equation \cite{exact,CCL98},
that under specific conditions
which could be realized at the moment when fireball starts to decay,
e.g., at the end of the pure hydrodynamic stage of the matter
evolution, spectra and interferometry radii may
become ``frozen'' at this early moment
despite the particle emission and collisions between hadrons last
for a fairly long period of the system decay.
It was also proposed \cite{Sinyukov} that this
peculiar theoretical result  can
be used to test numerical transport models of A+A collisions and
this is one of the aims of this paper.
Another aim is a study, within kinetic approach,
to what extent is the conservation of
spectra and interferometry radii violated during the system evolution
in the case of more realistic initial conditions.
This study is particularly interesting in the context of recently
observed \cite{akksin} approximate conservation of
interferometry volume during isentropic and chemically frozen
hydrodynamic evolution of hadron-resonance gas and, it can be
quite useful to solve the interferometry puzzle.

The analysis of above phenomena has been done within Universal Kinetic
Model (UKM) \cite{Am2002}, a Monte Carlo event generator,
realized as a numerical code written in object oriented C++ language.
The UKM gives a possibility to
model kinetic processes inside a particle system applying a suitable
numerical algorithm. This model is a universal one, since there is a
possibility to choose and add particle system (e.g., system of hadrons,
system of partons, etc.) and to choose and add a particular numerical
algorithm, i.e., to choose and add different kinetic modeler.
In our studies, we have used a cascade version of the UKM code,
UKM-R \cite{UKMRoot}, written under the ROOT system \cite{ROOT}.
We use the same numerical cascade algorithm as implemented in the
Quark Gluon String model \cite{AGTS90}.

\bigskip
The paper is now organized as follows.
In section~\ref{sec2} we relate particle spectra and
two--particle correlation function with distribution
and emission functions. Section~\ref{sec3} deals with the theoretical
aspects of the spectra
formation, including a possible duality between kinetic
and hydrodynamic approaches.
Section~\ref{sec4} is devoted to a brief UKM-R description.
In section~\ref{sec5} we discuss the known exact solution of the
nonrelativistic Boltzmann equation for an expanding fireball with
the spherically symmetric Gaussian initial conditions.
This solution is used in
section~\ref{sec6} to test the UKM-R.
Section~\ref{sec7} deals with the influence of
anisotropic initial conditions and relativistic effects
on the final spectra and interferometry radii.
The results are summarized in section~\ref{sec8}.

\section{\label{sec2}Distribution and emission functions.}

The inclusive particle spectra are related to the averages of the
products of creation and annihilation operators, $a_{p}^{+}$
and $a_{p}$, calculated at sufficiently large time
$t_{\mbox{\scriptsize out}}$ or, generally, on
an asymptotic hypersurface
$\sigma_{\mbox{\scriptsize out}}: t=
t_{\mbox{\scriptsize out}}(\mathbf{x})$,
guaranteeing the absence of further interactions.
Thus
\begin{equation}
p^{0}\frac{d^3N}{d^3\mathbf{p}}\equiv n(p)
=\langle a_{p}^{+}a_{p}\rangle ,~~~~~
p_{1}^{0}p_{2}^{0}\frac{d^6N}{d^3\mathbf{p\mathtt{_{1}}}
d^3\mathbf{p}\mathtt{_{2}}}\equiv n(p_1,p_2)
=\langle a_{p_{1}}^{+}a_{p_{2}}^{+}a_{p_{1}}a_{p_{2}}\rangle .
\label{spectra-def}
\end{equation}
In the case of independent (chaotic) particle production and
subsequent system kinetic
evolution, one can express the
average of the product of four operators through the irreducible
averages $\langle a_{p}^{+}a_{p'}\rangle$,
similar to Wick theorem for thermal systems.
For identical spin-0 bosons
\begin{equation}
 n(p_1,p_2)
=\langle a_{p_{1}}^{+}a_{p_{1}}\rangle
 \langle a_{p_{2}}^{+}a_{p_{2}}\rangle +
 \langle a_{p_{1}}^{+}a_{p_{2}}\rangle
 \langle a_{p_{2}}^{+}a_{p_{1}}\rangle.
\label{n12_spin-0}
\end{equation}
In kinetic approach,
the irreducible averages are expressed through the
integrals or
Fourier transforms of the distribution function $f(x,p)$
on the hypersurface $\sigma_{\mbox{\scriptsize out}}$:
\begin{equation}
\langle a_{p}^{+}a_{p}\rangle =
\int_{\sigma _{\mbox{\scriptsize out}}}d^3\sigma _{\mu
}(x)p^{\mu } f(x,p),~~~
\langle a_{p_{1}}^{+}a_{p_{2}}\rangle=
\int_{\sigma _{\mbox{\scriptsize out}}}d^3\sigma _{\mu
}(x)p^{\mu } e^{-iqx}f(x,p),
\label{average-wigner}
\end{equation}
where, in the nondiagonal average in the second integral,
$p=(p_{1}+p_{2})/2$ is an off-mass-shell four-momentum and
$q=p_{1}-p_{2}$.
The actual choice of
$t_{\mbox{\scriptsize out}}$ does not matter due to the
current conservation for free streaming particles
and, due to the fact that,
as a consequence of the relation $qp=0$, the four-vector
product $qx$ is independent of the position $x$
on the trajectory of a free off-mass-shell particle
with four-momentum $p$ (i.e., $qx$ is invariant with
respect to transformations $t\to t+\delta t$ and
$\mathbf{x}\to \mathbf{x}+(\mathbf{p}/p^0)\delta t$).

The distribution function $f(x,p)$
is closely related with
the current emission function $S^{\sigma}(\bar{x},p)$
on a hypersurface $\sigma: t=t({\bf x})$
since it
collects contributions of the emission or scattering points
$\bar{x}=\{\bar{t},\mathbf{x}-(\mathbf{p}/p^0)(t-\bar{t})\}$
starting from those free particles with the three-velocity
$\mathbf{p}/p^0$ reach a point $x=\{t,\mathbf{x}\}$.
Particularly, choosing the hypersurface $\sigma: t=const$,
one has
\begin{equation}
p^0f(x,p) =
\int d^4 \bar{x}\,\delta^3\left(\bar{\mathbf{x}}-
\mathbf{x}+(\mathbf{p}/p^0)(t-\bar{t})\right)
S^{\sigma}(\bar{x},p).
\label{f-S}
\end{equation}
Inserting Eq.~(\ref{f-S}) into Eq.~(\ref{average-wigner})
and using the equality $q\bar{x}=qx$,
one can rewrite the asymptotic operator
averages as four-volume integrals
or Fourier transforms of the asymptotic emission function
$S(\bar{x},p)\equiv S^{\sigma _{\mbox{\scriptsize out}}}(\bar{x},p)$
\cite{als02}:
\begin{equation}
\langle a_{p}^{+}a_{p}\rangle =
\int d^4 \bar{x} S(\bar{x},p),~~~
\langle a_{p_{1}}^{+}a_{p_{2}}\rangle =
\int d^4 \bar{x} e^{-iq\bar{x}}S(\bar{x},p).
\label{average-S}
\end{equation}

Using Eqs.~(\ref{spectra-def})-(\ref{average-S}),
one can write the two-boson correlation function (CF) as
\begin{equation}
{\cal R}(p_1,p_2)\equiv
\frac{n(p_1,p_2)}{n(p_1)n(p_2)}=
1+|{\cal F}(p,q))|^2 = 1+\langle \cos(qx_{12})\rangle',
\label{cf}
\end{equation}
where $x_{12}=x_1-x_2$,
\begin{equation}
{\cal F}(p,q) =\frac{
\langle a_{p_{1}}^{+}a_{p_{2}}\rangle}
{[\langle a_{p_{1}}^{+}a_{p_{1}}\rangle
\langle a_{p_{2}}^{+}a_{p_{2}}\rangle]^{1/2}}
\label{ff}
\end{equation}
and the quasiaverage
$$
\langle \cos(qx_{12})\rangle'=
\frac{\int d^{3}\sigma _{\mu }(x_{1})d^{3}\sigma
_{\nu }(x_{2})p^{\mu }p^{\nu }f(x_{1},p)f(x_{2},p)\cos (qx_{12})}{%
\int d^{3}\sigma _{\mu }(x_{1})d^{3}\sigma _{\nu }(x_{2})p_{1}^{\mu
}p_{2}^{\nu }f(x_{1},p_{1})f(x_{2},p_{2})}
$$
\begin{equation}
=
\frac{\int d^{4}\bar{x}_{1}d^{4}\bar{x}_{2}\, \cos(q\bar{x}_{12})\,
S\left(\bar{x}_{1},p\right) S\left(\bar{x}_{2},p\right) }
{\int d^{4}\bar{x}_1\, S\left(\bar{x}_1,p_1\right)
\int d^{4}\bar{x}_2\, S\left(\bar{x}_2,p_2\right)}.
\label{qa}
\end{equation}
Obviously, the CF decreases from 2 at $q=0$ to unity at large
$|q^\mu|$
and the width of this Bose-Einstein enhancement is inversely related
to the effective system size which is usually encoded in so called
interferometry radii.

In this paper, we will also
use Eqs.~(\ref{spectra-def})-(\ref{qa})
to calculate spectra and CF's at any earlier evolution times
$t_{\sigma} < t_{\sigma_{\mbox{\scriptsize out}}}$
on a hypersurface $\sigma$.
To do so, one has to make only the substitutions
$\sigma_{\mbox{\scriptsize out}} \to \sigma$ and
$S(\bar{x},p)\to S^{\sigma}(\bar{x},p)$.
Note that the current emission function
$S^{\sigma}(\bar{x},p)$ takes into account the kinetic
evolution up to the
evolution time $t_{\sigma}$ only and satisfies the obvious
relations: $S^{\sigma}(\bar{x},p)\ge S(\bar{x},p)$
for $\bar{t}\le t_{\sigma}$ and $S^{\sigma}(\bar{x},p)=0$ for
$\bar{t} > t_{\sigma}$.
Particularly, for a system of noninteracting
particles the emission function is independent
of the evolution time and the asymptotic one,
$S(\bar{x},p)$, coincides with that,
$S^{\sigma_0}(\bar{x},p)$,
determined on the initial hypersurface
$\sigma_0:~t=t_0(\mathbf{x})$.

{
To calculate the kinetic evolution,
we will use the simple ansatz
$S^{\sigma_0}(\bar{x},p)=p^0f(\bar{x},p)\delta(\bar{t})$
for the initial emission function on the hypersurface
$\sigma_0:~t=0$.
Though this ansatz generally violates the uncertainty relation
$\Delta \bar{t}\cdot\Delta p^0\ge 1$,
it causes no problem
since the evolution initial condition is
determined by the
initial distribution function only and
since there exist a whole family of the
initial emission functions of a finite time width
recovering the latter in accordance with Eq.~(\ref{f-S}).
}

\section{\label{sec3}Theoretical aspects of the spectra formation.}

The distribution function
describes an expanding system of interacting particles and
satisfies Boltzmann equation (BE) which, in the case of no
external forces, has the form:
\begin{equation}
{p^{\mu }}\frac{\partial f(x,p)}{\partial x^{\mu }}
=C_{\mbox{\scriptsize gain}}(x,p)-C_{\mbox{\scriptsize loss}}(x,p).
\label{Boltzmann}
\end{equation}
The collision terms $C_{\mbox{\scriptsize gain}}$ and
$C_{\mbox{\scriptsize loss}}$ are associated
with the numbers of
particles that respectively come to the phase space point $(x,p)$
and leave this point because of collisions.
Let us follow Ref. \cite{Sinyukov} and
split the distribution function at each
space-time point $x$ into two terms:
$f(x,p)=f_{\mbox{\scriptsize int}}(x,p)+f_{\mbox{\scriptsize esc}}(x,p)$.
The first one, $f_{\mbox{\scriptsize int}}(x,p)$,
describes the particles continuing to interact after the time $t$.
The second one, $f_{\mbox{\scriptsize esc}}(x,p)$,
describes the particles that do not interact any more after the time $t$.
Introducing escape probability $P(x,p)$, i.e.,
the probability that any \textit{given} particle at phase space
point $(x,p)$ does not interact
any more and propagates freely, one can write
\begin{equation}
f_{\mbox{\scriptsize esc}}(x,p)=P(x,p)\,f(x,p).  \label{prob-def}
\end{equation}
At large enough times
$t \sim t_{\mbox{\scriptsize out}}$,
$P(x,p)\rightarrow 1$ and
$f_{\mbox{\scriptsize esc}}(x,p)\rightarrow f(x,p)$.
Since in the evolution of $f_{\mbox{\scriptsize esc}}(x,p)$
there is no loss of particles
and the gain is associated with the part of the emission function,
$S_{\mbox{\scriptsize g}}(x,p)=[1-\theta(t_0-t)]S(x,p)$,
developed during
the kinetic evolution starting at a hypersurface $\sigma_0$
and describing particles suffering last
collisions at space-time point $x$, the escape function satisfies
the following BE \cite{Sinyukov}:
\begin{equation}
p^{\mu }\frac{\partial }{\partial x^{\mu }}
f_{\mbox{\scriptsize esc}}(x,p)=P(x,p)
C_{\mbox{\scriptsize gain}}(x,p)\,\equiv
S_{\mbox{\scriptsize g}}(x,p).  \label{eq-f+}
\end{equation}
Noting that
$\partial _{\mu }[p^{\mu }\exp{(-iqx)}]=0$
and that $f_{\mbox{\scriptsize esc}}(x,p)=f(x,p)$
on the asymptotic hypersurface ${\sigma_{\mbox{\scriptsize
out}}}$, one obtains,
applying the Gauss theorem to the Fourier transform
of the distribution function
in Eq.~(\ref{average-wigner}) and
using respectively general
equations (\ref{eq-f+}) and (\ref{Boltzmann})
analytically continued to off-mass-shell four-momenta $p$:
\begin{equation}
\langle a_{p_{1}}^{+}a_{p_{2}}\rangle =\int_{\sigma _{0}}d^3\sigma
_{\mu }(x_0)p^{\mu }f_{\mbox{\scriptsize esc}}(x_{0},p)
e^{{-iqx_0}}+\int_{\sigma _{0}}^{\sigma
_{\mbox{\scriptsize out}}}\!d^{4}x\,
S_{\mbox{\scriptsize g}}(x,p)e^{{-iqx}},
\label{sp-e}
\end{equation}
\begin{equation}
\langle a_{p_{1}}^{+}a_{p_{2}}\rangle =\int_{\sigma _{0}}d^3\sigma
_{\mu }(x_0)p^{\mu }f(x_{0},p)e^{{-iqx_0}}+\int_{\sigma _{0}}^{\sigma
_{\mbox{\scriptsize out}}}
d^{4}x\,\left[
C_{\mbox{\scriptsize gain}}(x,p)\!-\!C_{\mbox{\scriptsize
loss}}(x,p)\right]
e^{{-iqx}},
\label{sp-f}
\end{equation}
where $S_{\mbox{\scriptsize g}}(x,p)=P C_{\mbox{\scriptsize gain}}$
is defined in Eq. (\ref{eq-f+}),
$f_{\mbox{\scriptsize esc}}(x_{0},p)$ corresponds to the portion of the
particles that are
already free at initial time $t_{0}$
(generally, at initial hypersurface $\sigma _{0}$)
and $f(x_{0},p)$ is the distribution function
at $\sigma _{0}$.
Thus, as stated in Eqs.~(\ref{average-wigner}) and
(\ref{average-S}),
the Fourier transform of the distribution
or escape function on the asymptotic hypersurface
$\sigma_{\mbox{\scriptsize out}}$
is equivalent to the four-volume Fourier transform of the
emission function,
the latter being represented by the emission
function
$S_{\mbox{\scriptsize g}}=P C_{\mbox{\scriptsize gain}}$
developed during the kinetic evolution starting from
the initial hypersurface $\sigma _{0}$,
together with the
emission function
$\theta(t_0-\bar{t})S(\bar{x},p)$
corresponding to emission times $\bar{t}\le t_{\sigma _{0}}$.
The latter determines the escape function
$f_{\mbox{\scriptsize esc}}(x_0,p)$ on the hypersurface
$\sigma _{0}$
through
Eq.~(\ref{f-S}) modified by the substitutions
$f\to f_{\mbox{\scriptsize esc}},
~S^{\sigma}\to \theta(t_0-\bar{t})S(\bar{x},p)$.

One may see from Eq.~(\ref{sp-f}) that for a quasifree system
characterized by the equality
$C_{\mbox{\scriptsize gain}}=C_{\mbox{\scriptsize loss}}$,
the spectra and correlations are not affected by the kinetic
evolution and are completely determined by the initial
distribution function. An example is the system of nonrelativistic
particles with the initial spherically symmetric Gaussian
distribution function \cite{Sinyukov,CCL98}.

In hydrodynamic approach, the Landau freeze-out criterion
of locally equilibrium (leq) hydrodynamic momentum spectra
and corresponding Cooper-Frye prescription, defined by
Eq.~(\ref{average-wigner}) with substitutions
$\sigma _{\mbox{\scriptsize out}}\!\rightarrow
\sigma_{\mbox{\scriptsize f}}$
and $f\!\rightarrow f_{\mbox{\scriptsize leq}}$,
treats particle spectra as a result
of rapid conversion of a hadron system in local equilibrium
into a gas of free particles on
some freeze-out hypersurface $\sigma _{\mbox{\scriptsize f}}$.
Formally, it corresponds to taking the
cross section tending to infinity at
$t<t_{\sigma _{\mbox{\scriptsize f}}}$ (to
keep the system in local equilibrium) and zero beyond
$t_{\sigma _{\mbox{\scriptsize f}}}$. Then
$C_{\mbox{\scriptsize gain}}= C_{\mbox{\scriptsize loss}}$,
$P(x,p)=\theta (t-t_{\sigma _{\mbox{\scriptsize f}}})$,
$f_{\mbox{\scriptsize esc}}=
\theta (t-t_{\sigma _{\mbox{\scriptsize f}}})
f_{\mbox{\scriptsize leq}}$
and
$S=p^{0}
\delta (t-t_{\sigma _{\mbox{\scriptsize f}}})
f_{\mbox{\scriptsize leq}}$.
However, as stressed in Ref. \cite{Borysova},
in the \textit{realistic}
case of continuous freeze-out the emission function $S(x,p)$
is no more proportional to the distribution
function $f(x,p)$.
At the same time, the spectra and the interferometry radii
can be simply determined using the Cooper-Frye prescription
for the fireball characterized by the thermal distribution function
at the moment \textit{before} it starts to emit
particles and decays.
As analyzed in detail in Ref.~\cite{Sinyukov},
such a duality between hydrodynamic
and transport approaches
can happen when the integral of the difference
$C_{\mbox{\scriptsize gain}}-C_{\mbox{\scriptsize loss}}$
in Eq.~(\ref{sp-f})
vanishes, as is the case for a system of nonrelativistic
particles with the initial spherically symmetric
Gaussian distribution function (see sections~\ref{sec5} and \ref{sec6}),
or, when this integral is small, what can happen
for a more wide class of initial conditions (see section~\ref{sec7}).

\maketitle
\section{\label{sec4} Universal kinetic model.}

In our studies, we have used
the UKM-R code \cite{UKMRoot} written under ROOT
system. In this code one deals with particle objects and their lists.
Any particle object has its static and dynamic attributes. The
static attributes are such as particle names, charges, masses or
widths.
The dynamic attributes are particle four-coordinates and four-momenta.
The particle objects are grouped into particle lists.
The ROOT classes are used to describe particle properties, to
operate with three- and four-vectors and to organize containers.

\bigskip
In the present work, we have considered only one type of particles
and taken into account only their elastic binary collisions.
We have thus created
a specific list of {\it primary} particles
specifying their
spatial and momentum distributions and derived the
{\it Elastic interaction} and {\it Cascade model} classes from
the respective UKM
{\it Particle interaction} and {\it Universal kinetic algorithm} ones.
The {\it Cascade model} class represents the BE solver.
The used UKM-R deals with the following lists:

(i) {\it Primaries},
(ii) {\it Collisions},
(iii) {\it Secondaries},
(iv) {\it New secondaries}\\
and the following basic classes:

(i) {\it Initial state},
(ii) {\it Elastic interaction},
(iii) {\it Cascade model}.

\bigskip
{The numerical algorithm \cite{AGTS90} can be described as follows:}

\begin{enumerate}
\item  The list of {\it primaries} is
initialized and an initial value of the current time is set.

\item At first, all possible pairs of colliding
particles from the {\it primaries} list
are searched for and the {\it collision} list is filled in
according to the increasing collision time.
The binary particle collision is considered as possible
whenever the distance $d$ of closest approach of the two
particles in their c.m. system is less than the collision distance
calculated from the total collision cross section
which is a function of the two--particle c.m.
energy $\sqrt{s}$, i.e.,
\begin{equation}
d<\left[{\sigma \left(\sqrt{s}\right)/{\pi }}\right]^{1/2}.
\label{B5}
\end{equation}
In the present study, only elastic scattering is considered
and a constant test cross section is used.

\item  The earliest collision is simulated,
the two collided particles are removed from
all lists and the created particles are put in
the list of {\it new secondaries}.
The particle coordinates are shifted in accordance with the
time of simulated collision and particle velocity vectors.

\item  The collisions of the particles from
{\it new secondaries} list
with all particles from {\it primaries} and
{\it secondaries} lists are searched for and the
time ordered {\it collision} list is updated.

\item  The content of {\it new secondaries} list is moved to
{\it secondaries} list and simulation continues starting from
step 3 or stops when the {\it collision} list is empty
or the current time exceeds the user defined value
of the stop time.
\end{enumerate}

It should be noted that the above algorithm violates
the Lorentz covariance \cite{Covariance98}
since the condition $(\ref{B5})$
introduces action at distance and breaks local character of
the BE (see Eqs.~(\ref{Boltzmann}) and (\ref{T2})).
One can use the invariance of the BE with respect to
the transformation
$f\rightarrow kf,~
\sigma \rightarrow \sigma /k$
and diminish nonlocality of the algorithm
by simultaneously increasing the number of
particles and the inverse interaction cross
section by the same factor $k$.

\maketitle
\section{\label{sec5} Exact solution of nonrelativistic BE.}

Let us consider the kinetic evolution of a
system of $N$ nonrelativistic identical spin-0 particles of
mass $m$ assuming the initial conditions corresponding to
the global thermal Boltzmann distribution with temperature
$T_{0}$ (i.e., the absent initial flow) and
spherically symmetric Gaussian profile of
particle density with radius $R_{0}$. Assuming further
only elastic binary collisions in the system,
the collision integral in BE
(\ref{Boltzmann}) takes the form
\begin{equation}
C_{\mbox{\scriptsize gain}}(x,p)-
C_{\mbox{\scriptsize loss}}(x,p)
=\int
\frac{d^{3}\mathbf{p}_{1}
}{2p_{1}^{0}}d\Omega
[f(x,p')f(x,p_1')-f(x,p)f(x,p_1)]
F(s)\sigma (\sqrt{s},\theta ),  \label{T2}
\end{equation}
where the particle four-momenta in the binary scattering
satisfy the conservation law:
$p+p_1=p^{\prime}+p^{\prime}_1$,
$d\Omega=d\cos\theta d\phi$, $\cos\theta=1+2t/(s-4m^2)$,
$s=(p+p_1)^2$, $t=(p-p')^2$,
$F(s)=\frac{1}{2}[s(s-4m^{2})]^{1/2}$
is the invariant flow of two colliding particles of mass $m$ at
c.m. energy $\sqrt{s}$ and
$\sigma (\sqrt{s},\theta )={d\sigma _{el}}/{d\Omega }$
is the differential cross section of the elastic scattering
on the c.m. angle $\theta $.
Then, in the nonrelativistic case,
the BE has the following analytical solution for the
distribution function $f(x,p)\equiv f(t,\mathbf{r},\mathbf{p})$
\cite{exact}:
\begin{equation}
f(t,\mathbf{r},\mathbf{p})=\frac{N}{
\left[2\pi R_{0} (mT_{0})^{1/2}\right]^{3}}\,
\exp\left(-\frac{(\mathbf{r}-t\mathbf{p}/m)^{2}}{2R_{0}^{2}}
-\frac{\mathbf{p}^{2}}{2mT_{0}}\right).
\label{T3}
\end{equation}
Since Eq.~(\ref{T3}) represents also special case of the
solution of hydrodynamic equations for ideal fluid with the
radial initial flow,
it can be rewritten in the local equilibrium form \cite{CCL98}:
\begin{equation}
f(t,\mathbf{r},\mathbf{p})=
f_{\mbox{\scriptsize leq}}=
\frac{N}{
\left[2\pi R_{0} (mT_{0})^{1/2}\right]^{3}}\,
\exp \left(-\frac{\mathbf{r}^{2}}{2R^{2}(t)}
-\frac{(\mathbf{p}-m\mathbf{u}(t,\mathbf{r}))^{2}}
{2mT(t)}\right),  \label{T4}
\end{equation}
where $T(t)=T_{0}/\varphi(t)$, $R^2(t)=R_0^2 \varphi(t)$ and
$\mathbf{u}(t,\mathbf{r})=\mathbf{r} \dot{\varphi}(t)/\varphi(t)=
\mathbf{r} t T(t)/(mR_{0}^{2})$
are respectively the temperature, the system Gaussian radius
squared and the flow velocity,
$\varphi(t)=1+T_{0}t^{2}/(mR_{0}^{2})$;
the corresponding chemical potential
$\mu(t,r)=-\frac12 T(t) \mathbf{r}^{2}/R^2(t)$.

Inserting solution (\ref{T3}) into the BE, one can check that
its l.h.s. vanishes,
$
m{\partial f}/{\partial t}+
{\mathbf{p}}{\partial f}/{\partial\mathbf{r}}=0,
$
so the collision integral (its r.h.s.) should also vanish.
Indeed, inserting the solution (\ref{T4}) into
the factor in square brackets in Eq.~(\ref{T2}),
one can see that the collision integral vanishes (irrespective
of energy and angular dependence of the elastic cross section)
as a consequence of the energy and three-momentum
conservation:
\begin{eqnarray}
[f(x,p')f(x,p_1')-f(x,p)f(x,p_1)]
&\propto&
\exp \left(-\frac{\mathbf{p}^{\prime 2}+
\mathbf{p}_{1}^{\prime 2}}{2mT(t)}+
\frac{\mathbf{p}^{\prime }+\mathbf{p}_{1}^{\prime }}{T(t)}
\mathbf{u}(t,\mathbf{r})\right)
\nonumber \\
&-&
\exp \left(-\frac{\mathbf{p}^{2}+\mathbf{p}_{1}^{2}}{2mT(t)}+
\frac{\mathbf{p}+\mathbf{p}_{1}}{T(t)}\mathbf{u}(t,\mathbf{r})
\right)=0.
\label{T6}
\end{eqnarray}
This solution thus corresponds to a quasifree motion and so,
as discussed in section~\ref{sec3} (see, e.g., Eq.~(\ref{sp-f})),
the particle spectra and correlations are not
affected by the kinetic evolution and
stay unchanged at any times \cite{Sinyukov,CCL98}.
One may see from Eqs.~(\ref{T3}) and (\ref{T4}) that the
conservation of the momentum spectrum is related to the fact that
the decreasing temperature is compensated by increasing
flow velocity. The latter also compensates the effect
of increasing system size on the two-boson CF
thus guaranteeing the conserved interferometry radius.
Indeed, using distribution function (\ref{T3}),
explicit calculations yield time-independent momentum spectrum
and correlation function:
\begin{equation}
\frac{d^3N(t;\mathbf{p})}{d^3\mathbf{p}}=\frac{N}
{\left(2\pi mT_{0}\right)^{3/2}}
\exp \left(-\frac{
\mathbf{p}^{2}}{2mT_{0}}\right),  \label{np}
\end{equation}
\begin{equation}
{\cal R}(t;\mathbf{p}_1,\mathbf{p}_2)=
1+\exp \left[-\left(R_0^2-\frac{1}{4mT_{0}}
\right)\mathbf{q}^2\right],  \label{cfg}
\end{equation}
where $(4mT_0)^{-1/2}\equiv R_{\mbox{\scriptsize th}}$
measures the characteristic
thermal size (heat de Broglie length) of the single-pion
emitter.
At the same time, as seen from Eq.~(\ref{T4}),
the system expands and spatial density decreases with increasing
time:
\begin{equation}
\frac{d^3N(t;\mathbf{r})}{d^3\mathbf{r}}=\frac{N}
{\left(2\pi R^2(t)\right)^{3/2}}\!
\exp \left(-\frac{\mathbf{r}^{2}}{2R^{2}(t)}
\right),  \label{nr}
\end{equation}
where $R^{2}(t)=R_0^2+t^2T_0/m$.

\maketitle
\section{\label{sec6} Test of the UKM-R algorithm.}

\begin{figure}
\includegraphics[width=14cm]{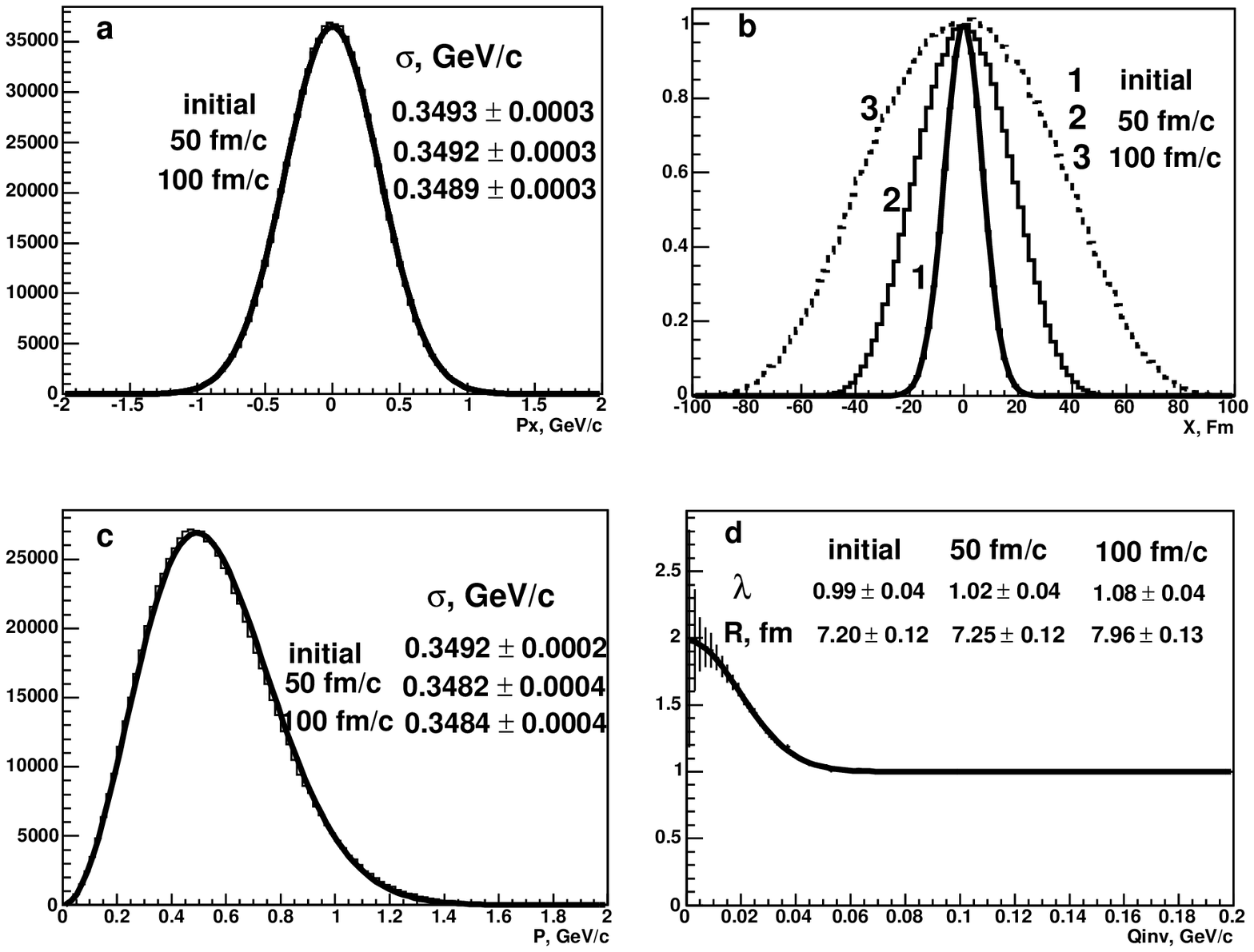}
\caption{\label{fig1} 
Distributions of the $x$-components of particle
three-momentum ({\bf a}) and three-coordinate ({\bf b}),
particle momentum ({\bf c}) and the corresponding CF's as functions
of $q_{inv}$ ({\bf d}) obtained in the UKM-R simulation of the
kinetic evolution of $N=400$ heavy spin-0 bosons of mass
$m=0.938$ GeV/c$^2$ at the
evolution time $t=0$, 50 and 100~fm/c.
The elastic cross section $\sigma_{el}=400$~mb, the
initial Gaussian radius $R_0=7$ fm and the initial
temperature $T_0=0.130$~GeV. The shown results of Gaussian fits
of the momentum distributions and the CF's agree
with the input initial values $\sigma=(mT_0)^{1/2}=0.3492$~GeV/c
and $\lambda=1$, $R_0=7$~fm respectively.
}
\end{figure}

To test the UKM-R algorithm, we have implemented the above
spherically symmetric Gaussian initial conditions and simulated
the kinetic evolution in a number of events
with fixed boson multiplicity $N=400$, choosing
$T_{0}=0.130$~GeV and $R_{0}=7$~fm. To guarantee the
nonrelativistic particle momenta, we have put $m=0.938$~GeV/c$^2$.
The initial momentum and coordinate distributions are shown
in Fig.~\ref{fig1}.
We have done two sets of simulations with
the elastic cross sections
$\sigma_{el} =$ $40$ and $400~$mb,
assuming the isotropic scattering in the two-particle c.m. system.
We have checked that the final results remain unchanged when
introducing the cross section anisotropy.

Instead of the six-dimensional CF in Eq.~(\ref{cf}),
we have calculated one-dimensional ones integrating
in numerator and denominator over all kinematic variables
except a chosen one, e.g.,
the modulus ${\rm q}=|\mathbf{q}|$ of the relative momentum vector
or its projections, or, the invariant relative momentum
$q_{inv}=(\mathbf{q}^2-q_0^2)^{1/2}=
\mathrm{q}(1-v_q^2)^{1/2}$,
where $v_q$ is the projection of the pair velocity vector
on the $\mathbf{q}$-direction; for nonrelativistic
particles $v_q^2\ll 1$ and $q_{inv}\doteq {\rm q}$.
As usual, we have assumed sufficiently smooth behaviour of the
single-particle momentum spectrum (i.e., in our case, the
initial thermal radius $R_{\mbox{\scriptsize th}}$
much smaller than the initial
system radius $R_0$) and substituted the quasiaverages in
Eq.~(\ref{qa}) by the averages making
the respective substitutions
$x_i,p \to x_i,p_i$ and $\bar{x}_i,p \to \bar{x}_i,p_i$
in the arguments of the
distribution and emission functions in the numerators.
Explicit calculation at $t=0$
then leads to the Gaussian CF in Eq.~(\ref{cfg}) with the
substitution $R_0^2-R_{\mbox{\scriptsize th}}^2 \to R_0^2$.
It should be noted however that this approximation violates
the invariance of the four-vector product $qx$ on the trajectory
of a free streaming particle (the invariance is guaranteed
only for free streeming particle with the off-mass-shell
four-momentum $p=(p_1+p_2)/2$). As a result, the averages
with distribution and emission functions are no more equivalent,
the latter being preferable since it avoids free streaming
from the emission or collision point to a given hypersurface
$\sigma$.

Concerning the approximate averaging using the distribution
function, at large evolution times $t$ as compared
with the the mean emission time
$\langle\bar{t}\rangle$,
it leads to the additional oscillating factor
in the CF,
\begin{equation}
{\cal R}(t;\mathbf{p}_1,\mathbf{p}_2)\doteq
1+\exp \left(-R_0^2\mathbf{q}^2\right)
\cos\left(\frac{t}{
m} q_{inv}^2\right).
\label{cfg1}
\end{equation}
The oscillating factor invalidates the smoothness
approximation at very large times and, at moderate times,
it leads to an increase of the correlation radius, roughly,
\begin{equation}
R_0^2\to
R_0^2+\left(\frac{k t}{
m R_0}\right)^2,
\label{R_incr}
\end{equation}
where the factor $k\sim 1$.
For heavy particles
of mass $m=0.938$ GeV/c$^2$ and the initial system
radius $R_0=7$ fm,
this increase does not exceed $10\%$ provided
$t < 100$ fm/c.

Similarly, the approximate averaging using the emission
function at a current evolution time $t$ yields
\begin{equation}
{\cal R}(t;\mathbf{p}_1,\mathbf{p}_2) \doteq
1+\exp \left(-R_0^2\mathbf{q}^2\right)
\left\langle
\cos\left(\frac{\bar{t}_{12}}{
m} q_{inv}^2\right)
\right\rangle_{\bar{t}_{1,2}\le t},
\label{cfg2}
\end{equation}
where $\bar{t}_{12}=(\bar{t}_1+\bar{t}_2)/2$
and the averaging is done over the emission times
$\bar{t}_{1,2}\le t$. For large evolution times,
$t\sim t_{\mbox{\scriptsize out}}$, the distortion factor
in Eq.~(\ref{cfg2}) is independent of $t$ and its effect
on the CF is determined by
$\langle \bar{t}_{12}\rangle \ll t_{\mbox{\scriptsize out}}$.
It thus leads to only a slight increase of the
correlation radius, roughly determined by Eq.~(\ref{R_incr})
with the substitution $t\to \langle \bar{t}_{12}\rangle$.

Practically, we have calculated the one--dimensional CF
in the bins of a chosen variable $q_j$ (e.g., $q_{inv}$)
as
\begin{equation}
{\cal R}(q_{j}) = \frac{1}{\Delta N(q_{j})}
\sum_{i=1}^{\Delta N(q_{j})}
[1 + \cos(q^{(i)} x_{12}^{(i)})],
\label{cfs}
\end{equation}
where $\Delta N(q_{j})$ is the number of simulated pairs
in a given bin, consisting of the particles from the same
event only; i.e., each pair from the same event was attributed
the weight $[1 + \cos(q^{(i)} x_{12}^{(i)})]$
and the weighted
histogram of $q_{j}$ was produced.
To determine the corresponding interferometry radius $R_j$,
the constructed CF has been fitted as
\begin{equation}
{\cal R}(q_{j}) = 1+\lambda\exp(-R_j^2q_{j}^2).
\label{cff}
\end{equation}

To check the exact solutions for the momentum and coordinate
distributions in Eqs.~(\ref{np}) and (\ref{nr}) and the CF
in Eq.~(\ref{cfg}),
the particle coordinates and momenta have been evaluated at the
following values of the evolution time $t$: 0, 50 and 100~fm/c.
In Fig.~\ref{fig1} we show the results of the UKM-R simulations
corresponding to the elastic cross section $\sigma_{el}=400$~mb;
we have checked that they coincide with those obtained for
$\sigma_{el}=40$~mb.
These results confirm the independence of the
momentum spectrum and the CF on the evolution time despite
that a huge number of collisions happened during the evolution.
A slight
increase of the fitted interferometry radius
at $t=100$ fm/c is a consequence of the smoothness
approximation and agrees
with Eqs.~(\ref{cfg1}) and (\ref{R_incr}).
As for the coordinate distribution, panel {\bf b} of Fig.~\ref{fig1}
confirms the
increase of the system Gaussian radius with the evolution
time in accordance with the law $R^{2}(t)=R_0^2+t^2T_0/m$.
The results shown in Fig.~\ref{fig1} are in full correspondence
with the exact solution of the BE for distribution function
and thus confirm the correctness of the kinetic simulation.

\begin{figure}
\includegraphics[width=14cm]{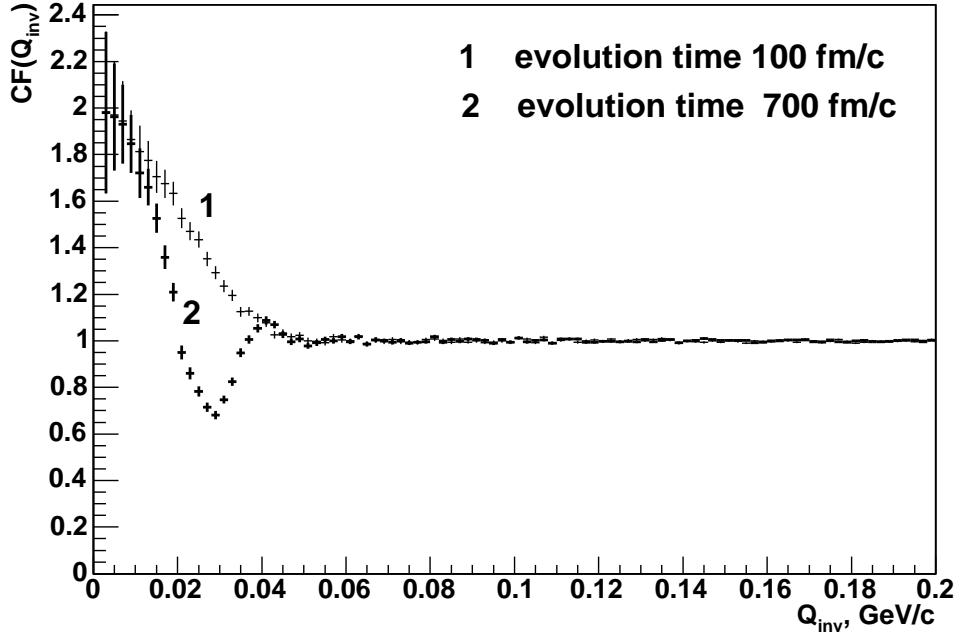}
\caption{\label{fig2} 
CF's obtained from four-coordinates of free streaming
($\sigma_{el}=0$) heavy spin-0 bosons of mass
$m=0.938$ GeV/c$^2$ obtained in the UKM-R simulation
on the same conditions as in Fig.~\ref{fig1}
at the evolution times $100$ and 700~fm/c.
 }
\end{figure}

\begin{figure}
\includegraphics[width=14cm]{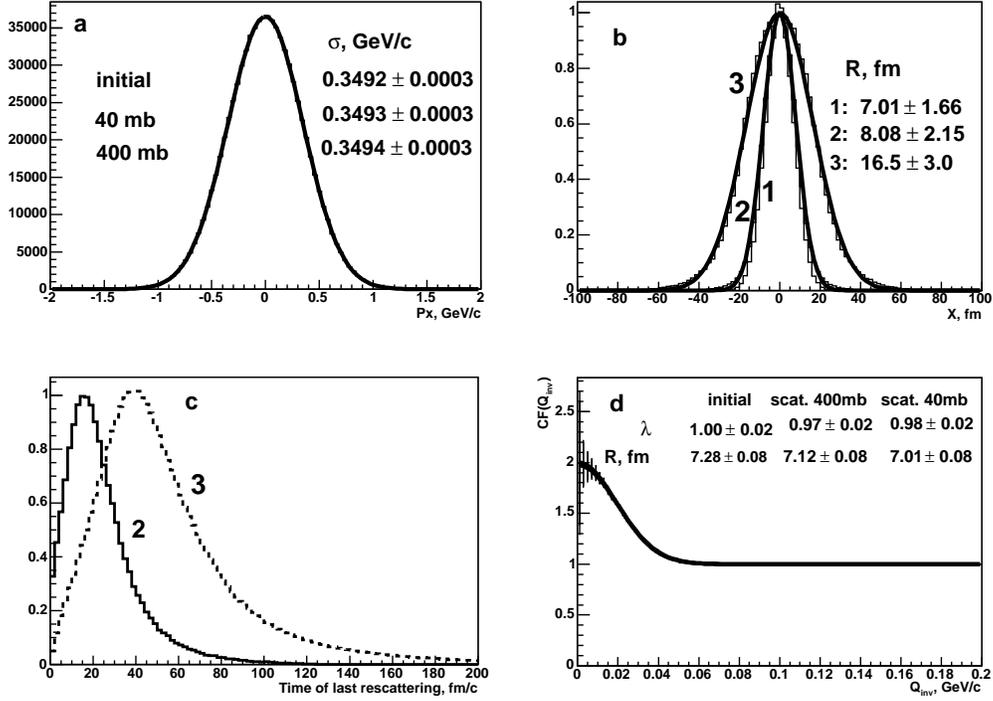}
\caption{\label{fig3} 
Initial and final distributions of the $x$-components of
particle three-momentum ({\bf a}),
spatial $x$- ({\bf b}) and time ({\bf c})
coordinates of the emission or last collision points
and the CF's corresponding to these points as functions of $q_{inv}$
({\bf d}) obtained
in the UKM-R simulation of $N=400$ heavy spin-0 bosons of mass
$m=0.938$ GeV/c$^2$ at the evolution time $t=$ 200 fm/c.
The histograms 1, 2 and 3 in panels {\bf b} and {\bf c}
correspond to
$\sigma_{el}=$ 0 (initial $x$-distribution), 40 and 400 mb
respectively.
The shown results of Gaussian fits
of the momentum distributions and the CF's agree
with the input initial values $\sigma=(mT_0)^{1/2}=0.3492$~GeV/c
and $\lambda=1$, $R_0=7$~fm respectively.
 }
\end{figure}

As noted after Eq.~(\ref{cfg1}), the use of the smoothness
approximation in the calculation of the CF with the help
of the distribution function is invalidated at large
evolution times.
This is demonstrated in Fig.~\ref{fig2}, where
the appearance of the artificial zero of
${\cal R}(q_{inv})-1$ at
$q_{inv}=(\pi m/2 t)^{1/2}\approx 0.02$ GeV/c
due to the oscillating factor in Eq.~(\ref{cfg1})
is clearly seen in the CF calculated from boson coordinates
at the evolution time $t=700$ fm/c.
To calculate the correct CF at asymptotic times,
we have therefore followed the system evolution
up to a sufficiently large time ($t\sim 200$~fm/c for
$\sigma_{el}=400$~mb)
and taken as the particle four-coordinates those of
the last emission or collision points. This corresponds to the
system description with the help of the asymptotic emission
function $S(x,p)$ which is much less affected by the smoothness
approximation as compared with the distribution function.
The spatial and time distributions of these points,
corresponding to $\sigma_{el}=40$ and 400~mb,
are shown in
the panels ${\bf b}$ and ${\bf c}$ of Fig.~\ref{fig3}.
For $\sigma_{el}=40$~mb, one may see only
a slight broadening of the
spatial distribution as compared with the initial one and a quite
narrow time distribution with a noticeable number of nonscattered
particles ($t=0$) and the asymptotic time of $\sim 100$~fm/c.
For $\sigma_{el}=400$~mb, the spatial and time distributions
are much broader,
almost all particles scatter at least once and the
asymptotic time is $\sim 200$~fm/c.
As for the asymptotic momentum spectra and interferometry radii,
one may see from the panels ${\bf a}$ and ${\bf d}$ of Fig.~\ref{fig3}
that they coincide with the initial ones within
the errors for both values of $\sigma_{el}$.

\maketitle
\section{\label{sec7} 
Anisotropic,
non-Gaussian and relativistic systems.}

We have also studied to what extent are the spectra and CF's
conserved in kinetic evolution of
nonrelativistic and relativistic systems characterized
by initial conditions
different from the spherically symmetric Gaussian ones
given in Eq.~(\ref{T3}) at $t=0$.

\subsection{\label{sec71} Nonrelativistic systems.}
\subsubsection{\label{sec711}Anisotropic non-Gaussian spatial distribution.}

\begin{figure}
\includegraphics[width=14cm]{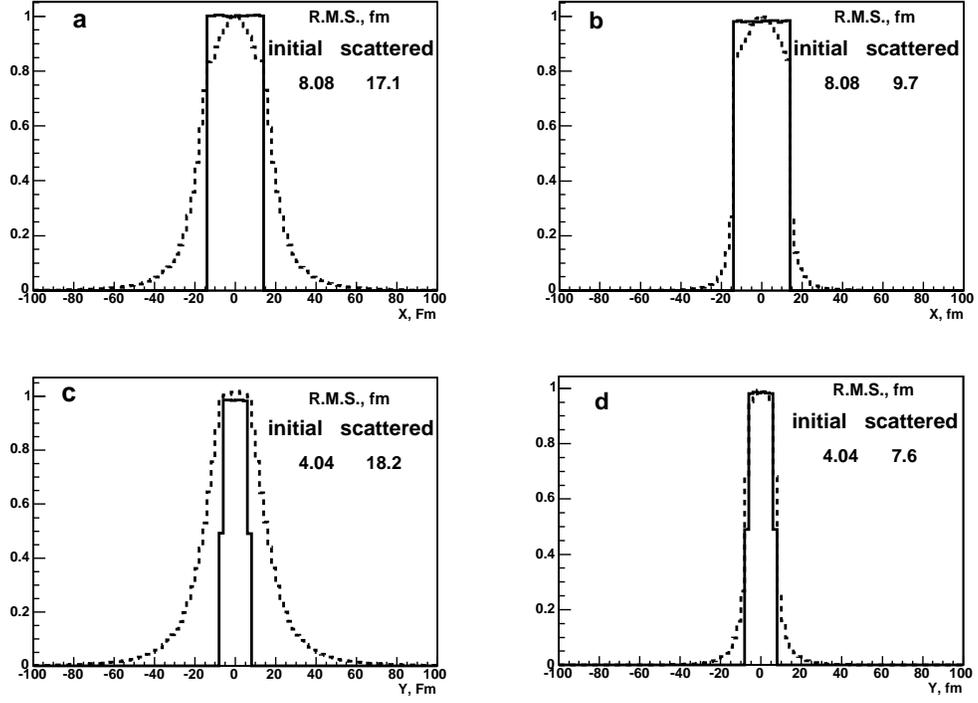}
\caption{\label{fig4} 
Final distributions of $x$- and $y$-coordinates of the emission
or last collision points (dashed histograms) obtained
in the UKM-R simulation on the same conditions as in Fig.~\ref{fig3}
except for the initial distributions of spatial coordinates
(solid histograms): $|x| \le 14$~Fm, $|y|, |z| \le 7$~Fm.
The panels {\bf a}, {\bf c}
and {\bf b}, {\bf d}
correspond to $\sigma_{el}=$ 400 and 40 mb respectively.
 }
\end{figure}

\begin{figure}
\includegraphics[width=14cm]{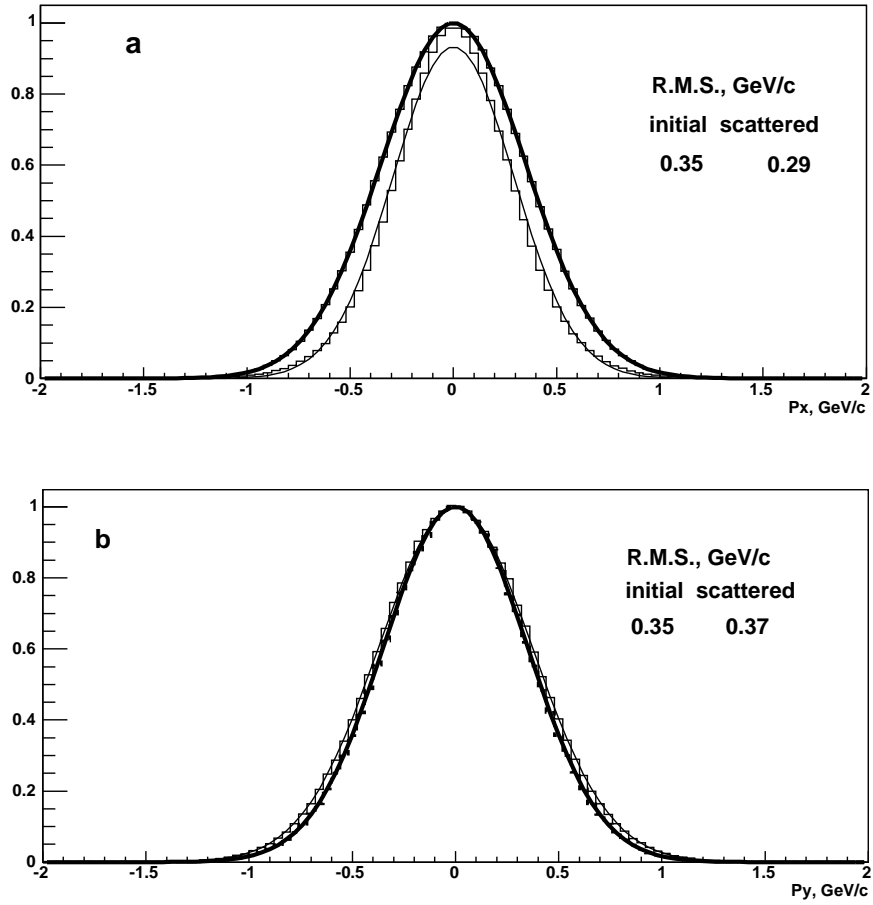}
\caption{\label{fig5}
Initial (thick) and final (thin) histograms of the components
of particle three--momentum
obtained in the UKM-R simulation on the same conditions
as in Fig.~\ref{fig4}.
 }
\end{figure}

\begin{figure}
\includegraphics[width=14cm]{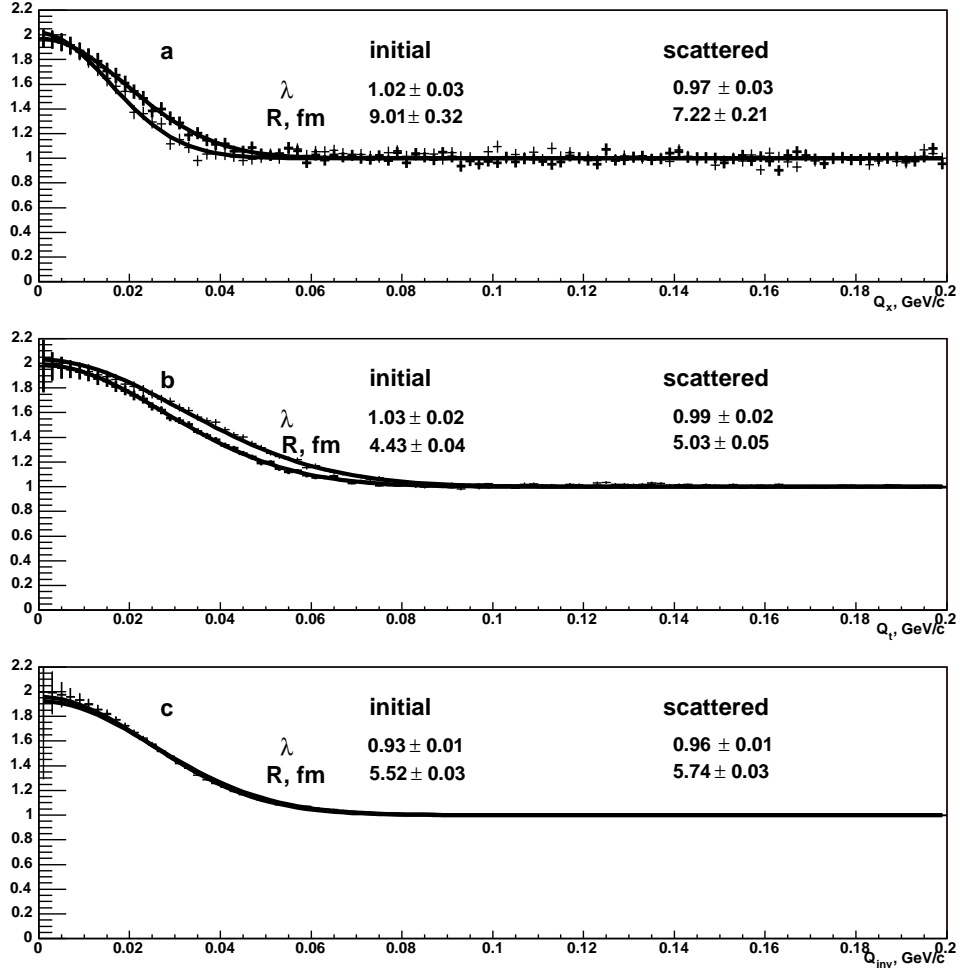}
\caption{\label{fig6}
 Initial and final CF's corresponding to the UKM-R
coordinate distributions in panels {\bf a} and {\bf c}
of Fig.~\ref{fig4} ($\sigma_{el}=400$ mb).
The CF's are shown as functions of $q_x$ ($q_t \le 10$~MeV/c)
({\bf a}),
$q_t = ({q_y}^2+{q_z}^2)^{1/2}$ ($q_x \le 7$~MeV/c) ({\bf b}) and
$q_{inv}$ ({\bf c}).
The results of Gaussian fits are shown. The fitted initial
Gaussian radii in panels {\bf a} and {\bf b}
are close to the r.m.s. values 8.08 and 4.04 fm
of the corresponding
initial $x$- and $y$-coordinate distributions
in panels {\bf a} and {\bf c} of Fig.~\ref{fig4}.
}
\end{figure}

As in section~\ref{sec6}, we have considered the nonrelativistic
system of $N=400$ heavy spin-0 bosons of
mass $m=0.938$ GeV/c$^2$ and the initial momentum distribution
given in Eq.~(\ref{np}) with $T_0=0.130$ GeV. However,
the initial spherically symmetric Gaussian spatial distribution
was substituted by a uniformly populated parallelepiped defined by
$|x|\le 14$~fm and $|y|, |z| \le 7$~fm (Fig.~\ref{fig4}); the corresponding
initial volume is about twice as large than in the previous case.
The kinetic evolution was performed with two values of the
elastic cross section, $\sigma_{el} =$ $40$ and $400~$mb.
In the former case, the collisions are quite rare
so the original spatial and momentum distributions,
as well as the CF, are practically not changed during the evolution.
In the latter case, the collisions are much more frequent and
force the spatial distribution of the emission or last
collision points to become close to a spherically symmetric Gaussian
one (see dashed histograms in panels {\bf a} and {\bf c}
of Fig.~\ref{fig4}).
On the other hand, as shown in Fig.~\ref{fig5},
the initial space anisotropy is transformed
in the anisotropy of the final three-momentum distribution.
As compared with the
originally spherically symmetric three-momentum distribution,
the final one
becomes softer in the $x$-direction and harder in the
$(y,z)$-plane.
As for the CF's, one may see from Fig.~\ref{fig6} that their evolution
reflects the one of the emission points,
the final CF's becoming closer to
spherical symmetry.
It is interesting to note that
despite of the noticeable difference of the initial and final
CF's, the interferometry volume,
$R_{x}R_{t}^{2}$, is practically not changed during
the evolution,
the initial and final one composing $177\pm 6$ and
$182\pm 6$ fm$^{3}$ respectively.
Also the $q_{inv}$ projection of the CF is changed only slightly
(see panel {\bf c} in Fig.~\ref{fig4}).
The effect of approximate
conservation of the interferometry volume
is in correspondence with a similar result
found for the chemically frozen
hydrodynamic evolution \cite{akksin}.

\subsubsection{\label{sec712}Nonthermal momentum distribution}

\begin{figure}
\includegraphics[width=14cm]{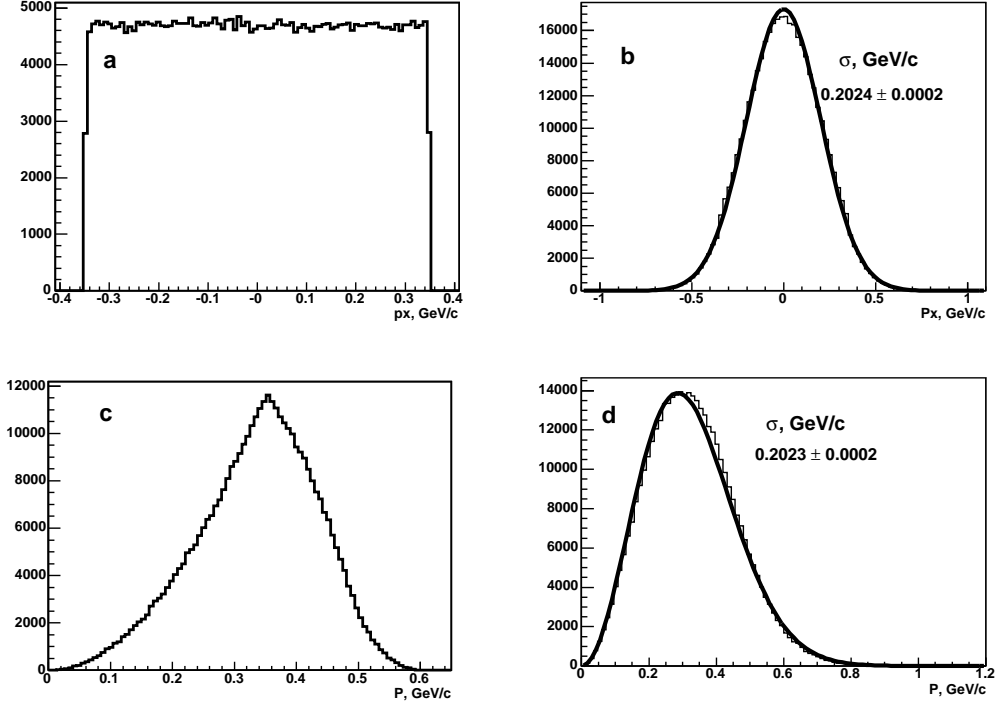}
\caption{\label{fig7}
Final distributions of the $x$-component ({\bf b}) and
the magnitude of particle
three--momentum ({\bf d}) obtained
in the UKM-R simulation with $\sigma_{el}=$ 400 mb
on the same conditions as in Fig.~\ref{fig3} except for the initial
three--momentum distribution ({\bf a} and {\bf c}) defined by
$|p_x|, |p_y|, |p_z| \le (T_{0}m)^{1/2}$,
$T_{0}$ = 0.130~GeV, $m=$0.938~GeV/c$^2$.
The fitted Gaussian widths of final distributions, shown in
panels {\bf b} and {\bf d}, are close to the r.m.s. value
$(mT_0/3)^{1/2}$ = 0.2016~GeV of the initial
$p_x$-distribution.
 }
\end{figure}

\begin{figure}
\includegraphics[width=14cm]{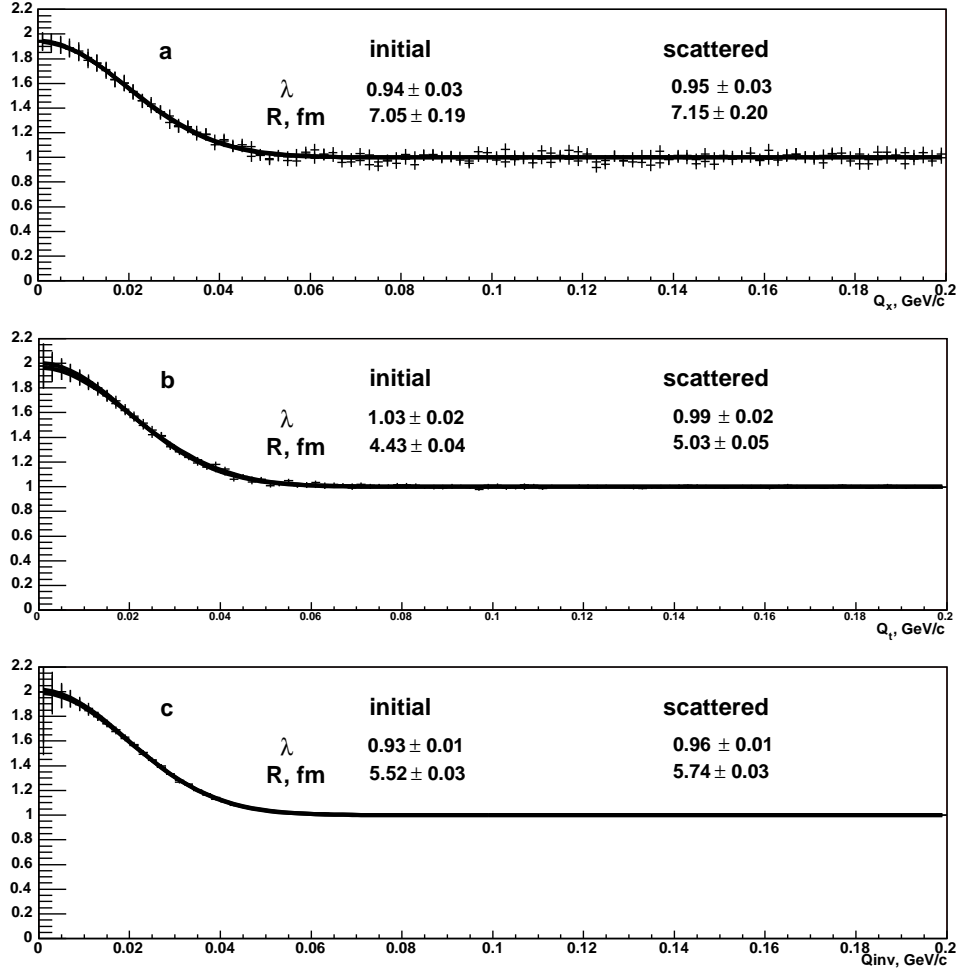}
\caption{\label{fig8}
Initial and final CF's corresponding to the UKM-R
simulation specified in the caption of Fig.~\ref{fig7}.
The CF's are shown as functions of $q_x$ ($q_t \le 10$~MeV/c)
({\bf a}),
$q_t = ({q_y}^2+{q_z}^2)^{1/2}$ ($q_x \le 7$~MeV/c) ({\bf b})
and $q_{inv}$ ({\bf c}).
The results of Gaussian fits are shown. The fitted
correlation radii
are close to the initial system Gaussian radius $R_0=7$~fm.
 }
\end{figure}

We have further considered the same initial conditions as in
section~\ref{sec6} except for the thermal Gaussian three--momentum
distribution which has been substituted by a uniformly populated
cube defined by $|p_x|, |p_y|, |p_z| \le (T_{0}m)^{1/2}$,
$T_{0}$ = 0.130~GeV, $m=$0.938~GeV/c$^2$ (Fig.~\ref{fig7}).
The results of the kinetic evolution of the momentum
distribution and the CF's corresponding to
the initial Gaussian radius $R_{0}=7$~fm
and the elastic cross section $\sigma_{el}= 400$ mb
are shown in figures \ref{fig7} and \ref{fig8}; similar results have been
obtained with $\sigma_{el}= 40$ mb.
One may see that the collisions force the initially
nonthermal distribution to take the thermal Gaussian form.
At the same time, r.m.s. of the initial distribution of the
momentum components,
$\sigma =(mT_{0}/3)^{1/2}=201.6$ MeV/c, as well as the initial
CF and the interferometry radius of 7 fm, are practically
conserved.
E.g., the Gaussian fits of the $p_x$ distribution and the
$q_{t}$ projection of the CF yield $\sigma =202.4\pm 0.2$ MeV/c
and $R_{t}=7.14\pm 0.11$ fm respectively.

\subsection{\label{sec72} Relativistic systems.}

\begin{figure}
\includegraphics[width=14cm]{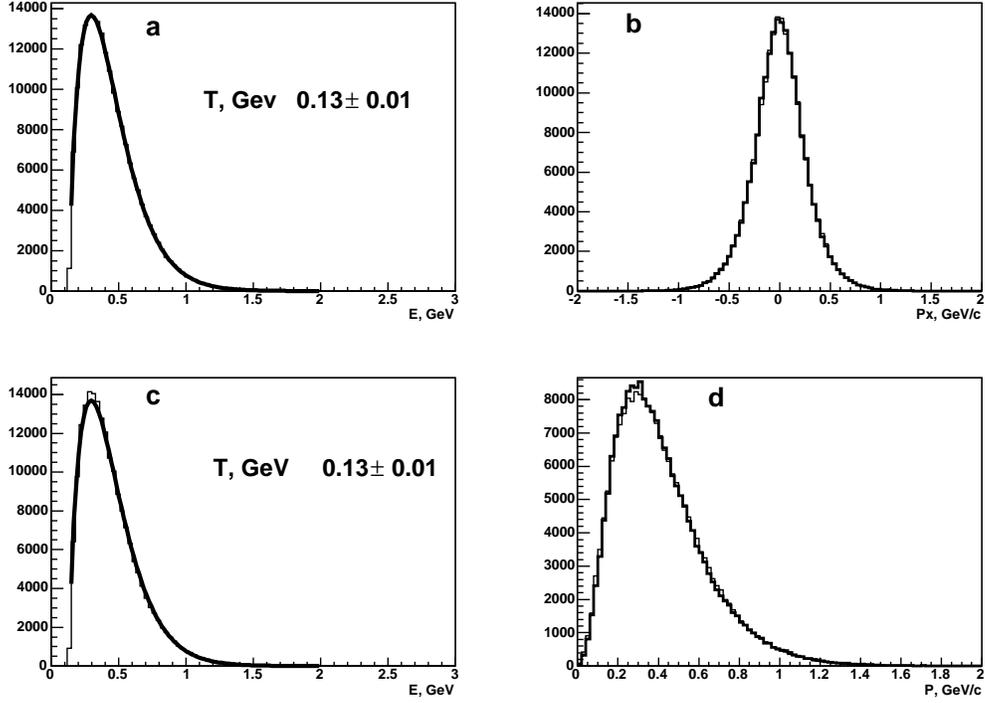}
\caption{\label{fig9}
Distributions of particle energy ({\bf a}, {\bf c}),
momentum ({\bf d}) and $x$-component of particle
three-momentum ({\bf b})
obtained in the UKM-R simulation of the
kinetic evolution of $N=400$ light spin-0 bosons of mass
$m=0.140$ GeV/c$^2$ and
elastic cross section $400$~mb.
The initial three--coordinates and three--momenta
have been respectively distributed according to
Gaussian of the radius $R_0=7$ fm and relativistic
thermal distribution with the temperature $T_0=0.130$~GeV.
The initial distributions are shown as solid histograms in
panels {\bf a}, {\bf b}, {\bf d} and the final ones -
as solid histogram in panel {\bf c} and dashed histograms
in panels {\bf b}, {\bf d}.
The fitted values of the initial and final temperature
shown in panels {\bf a} and {\bf c} coincide with $T_0$.
}
\end{figure}

\begin{figure}
\includegraphics[width=14cm]{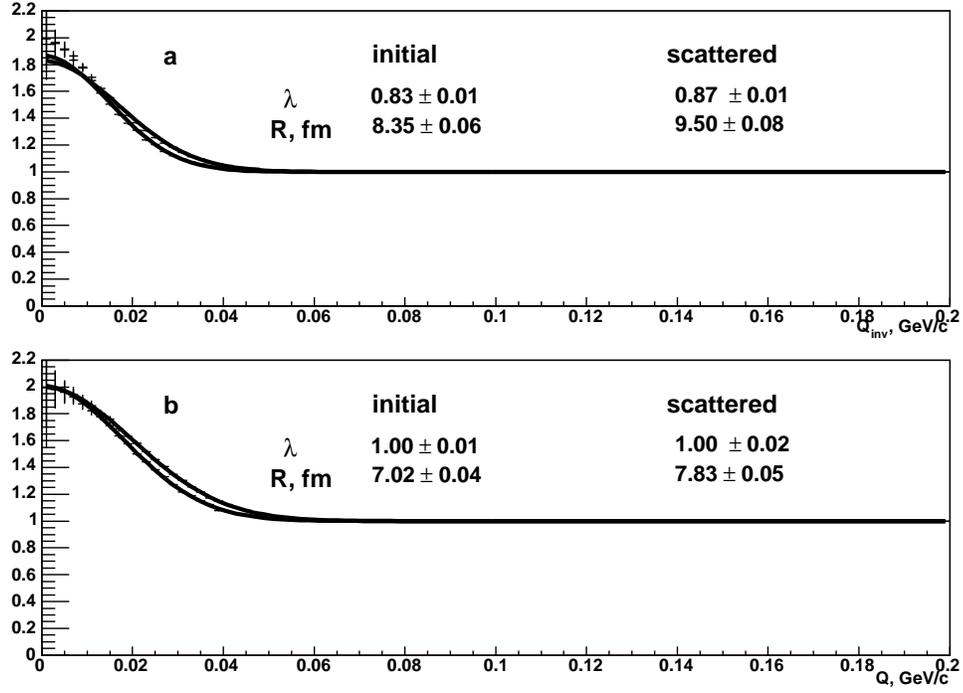}
\caption{\label{fig10}
Initial and final CF's corresponding to the UKM-R
simulation specified in the caption of Fig.~\ref{fig9}.
The CF's are shown as functions of $q_{inv}$ ({\bf a}) and
${\rm q}=|{\bf q}|$ ({\bf b}).
 }
\end{figure}

We have also studied the kinetic evolution of the
relativistic hadronic gas ($m \sim T_0$) choosing the
particle mass $m=0.140$~ GeV/c$^2$ and the initial
temperature $T_{0}=0.130$~GeV.
It is well known that the only relativistic solution of the
BE with vanishing collision term is the $x$-independent
J\"uttner distribution function,
\begin{equation}
f(x,p)=\frac{1}{(2\pi)^3}
\exp\left(\frac{\mu-p^\mu U_\mu}{T}\right),
\label{Juttner}
\end{equation}
corresponding to a global thermal equilibrium with
temperature $T$, chemical potential $\mu$ and system
four-velocity $U$.
Particularly,
the relativistic generalization of the distribution
function in Eq.~(\ref{T3}),
\begin{equation}
f(x,p)\propto
\exp\left(-\frac{(\mathbf{r}-t\mathbf{p}/p^0)^{2}}{2R_0^2}
-\frac{p^0}{T_{0}}\right),
\label{T3r}
\end{equation}
represents only an approximate solution of the BE,
leading to vanishing l.h.s.
but retaining a nonzero r.h.s. collision term.
Generally, one may expect that, irrespective of initial
conditions, the distribution function formed in the
kinetic evolution should tend to the J\"uttner one,
i.e., the spatial distribution should expand and the
three--momentum distribution $d^3N/d^3{\bf p}$
should tend to the thermal one
$\propto \exp[-({\bf p}^2+m^2)^{1/2}/T]$.
Particularly, starting the kinetic evolution from
the initial conditions corresponding to
Eq.~(\ref{T3r}) at $t=0$,
one may expect
an increase of the system Gaussian radius with the
evolution time and conserved momentum distribution.
Based on the analogy with the nonrelativistic system,
one can also expect an approximate conservation
of the correlation radius.
To check these expectations,
we have put the initial Gaussian
radius $R_{0} = 7$~fm and simulated the kinetic evolution
for $N=400$ particles with the elastic cross section
$\sigma_{el}=400$~mb.
The simulation results shown in figures~\ref{fig9} and ~\ref{fig10}
indeed confirm the conservation of the momentum
spectrum and approximate conservation of the CF,
the final correlation radius exceeding the initial one
by $\sim 9\%$ only.
We have obtained the same results for $N=40000$ particles and
$\sigma_{el}=4$~mb thus proving the negligible effect
of the violation of Lorentz covariance introduced by the
nonlocal condition in Eq.~(\ref{B5}); some deviations start
to appear only at $\sigma_{el} \sim 1000$~mb.

\subsubsection{\label{sec721}Anisotropic non-Gaussian spatial distribution.}

\begin{figure}
\includegraphics[width=14cm]{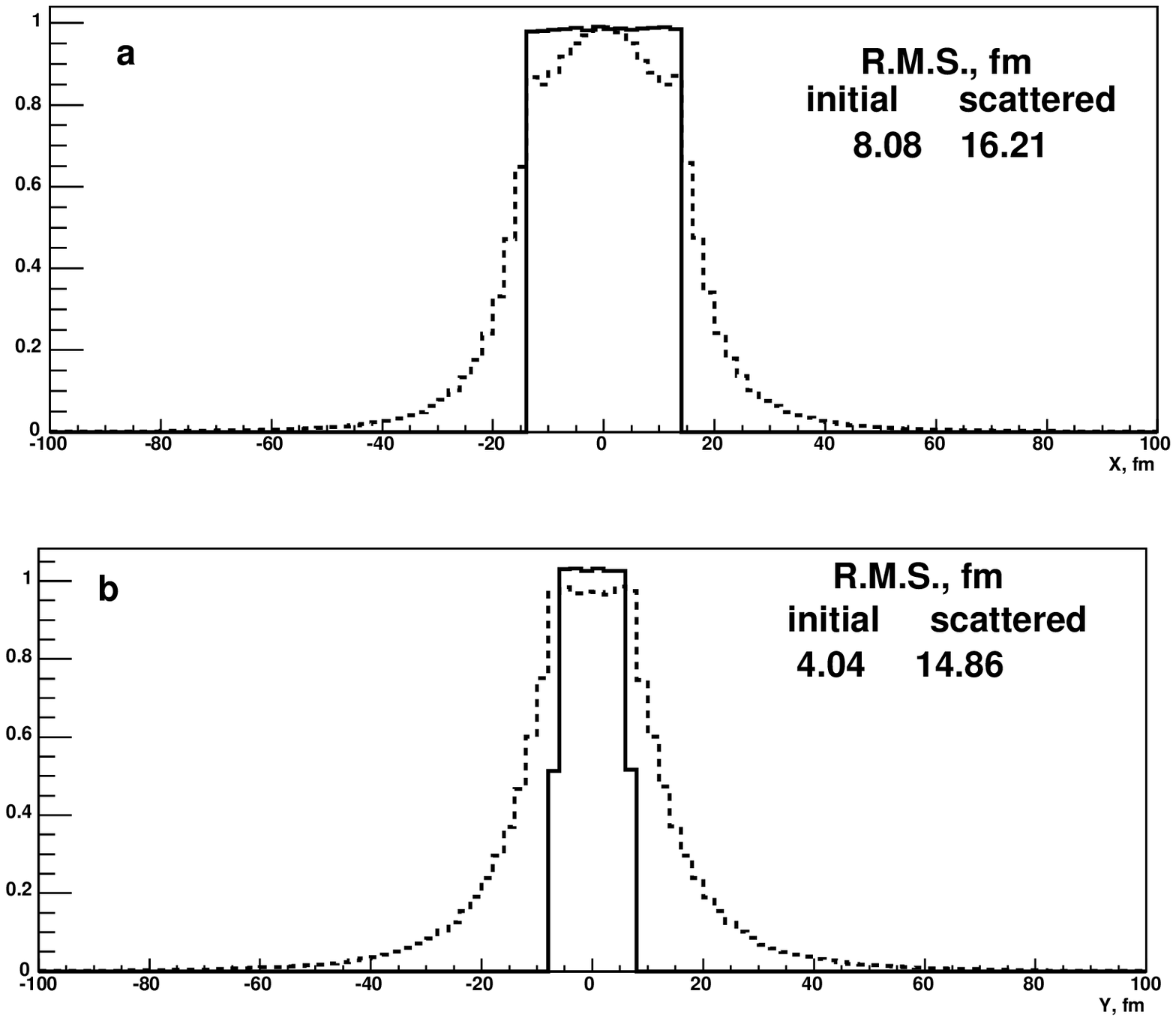}
\caption{\label{fig11}
Final distributions of $x$- ({\bf a}) and $y$- ({\bf b})
coordinates of the emission
or last collision points
(dashed histograms) obtained
in the UKM-R simulation on the same conditions as in Fig.~\ref{fig9}
except for the initial distributions of spatial coordinates
(solid histograms): $|x| \le 14$~Fm, $|y|, |z| \le 7$~Fm.
 }
\end{figure}

\begin{figure}
\includegraphics[width=14cm]{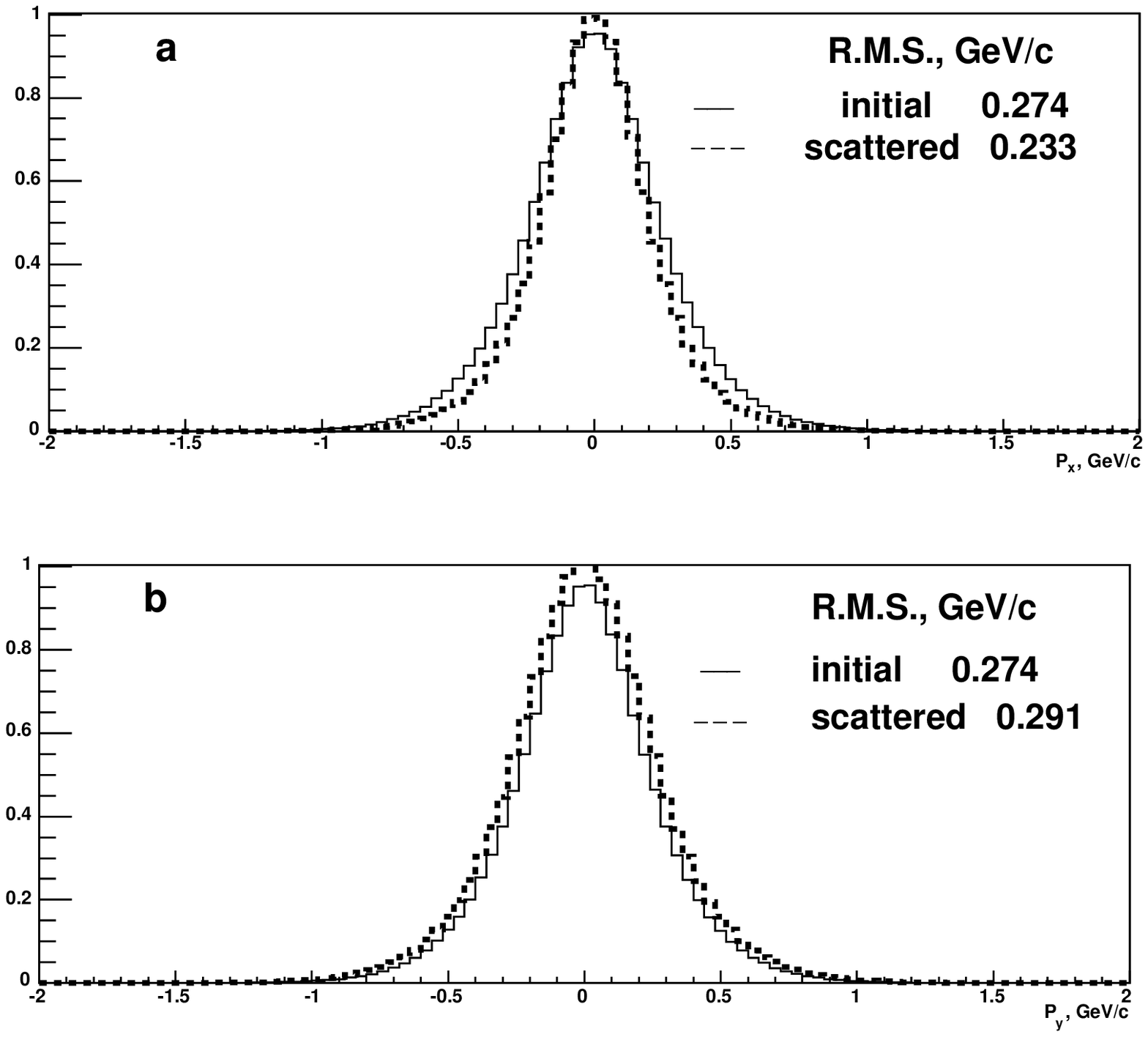}
\caption{\label{fig12}
Initial (solid) and final (dashed) histograms of the
components of particle three--momentum
obtained in the UKM-R simulation on the same conditions
as in Fig.~\ref{fig11}.
 }
\end{figure}

\begin{figure}
\includegraphics[width=14cm]{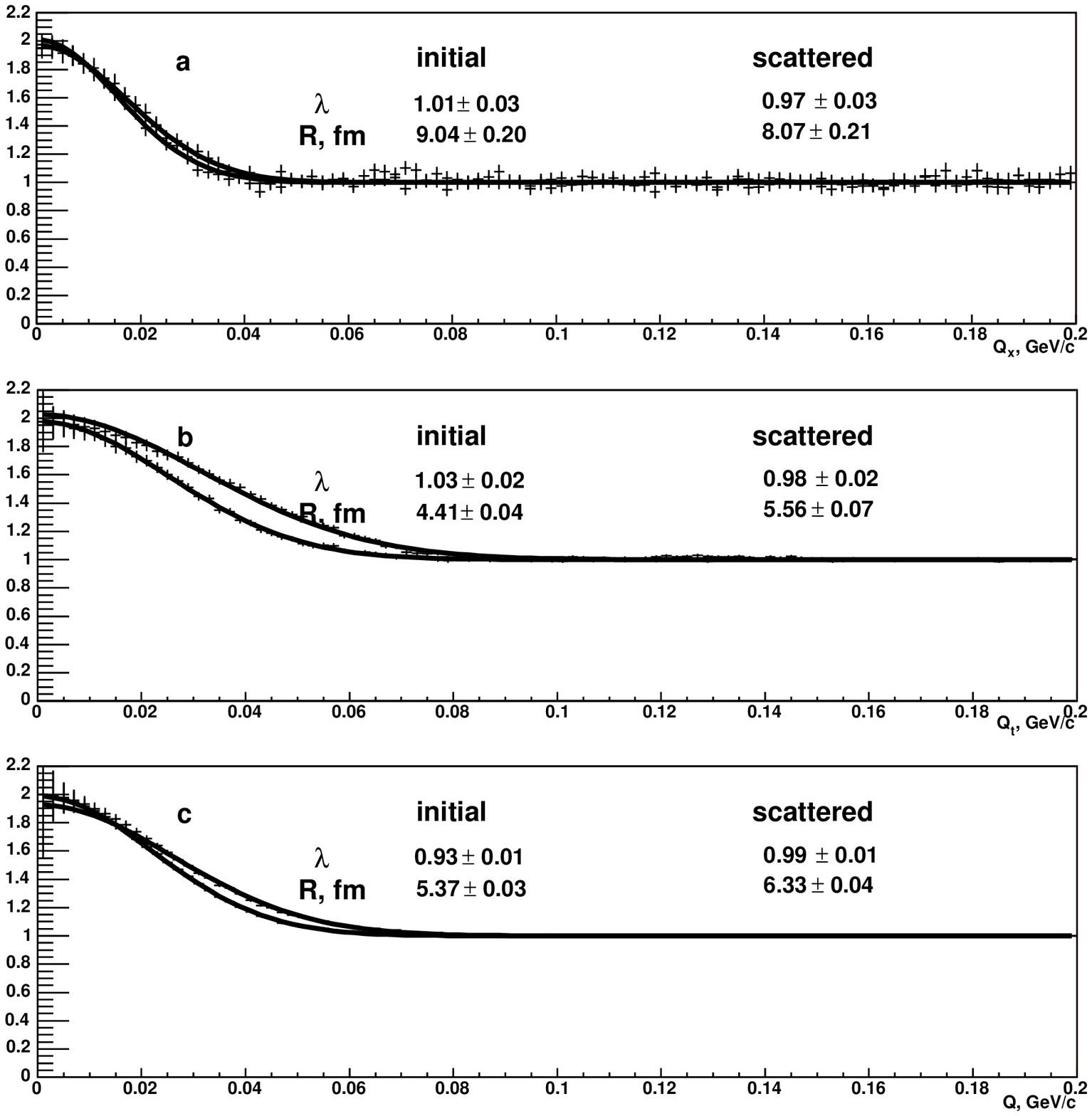}
\caption{\label{fig13}
Initial and final CF's corresponding to the UKM-R
simulation specified in the caption of Fig.~\ref{fig11}.
The CF's are shown as functions of
$q_x$ ($q_t \le 10$~MeV/c),
$q_t = ({q_y}^2+{q_z}^2)^{1/2}$ ($q_x \le 7$~MeV/c)
and ${\rm q}=|{\bf q}|$.
 }
\end{figure}

We have considered the same initial anisotropic
coordinate distributions as for the nonrelativistic bosons.
Comparing figures \ref{fig4}-\ref{fig6} and \ref{fig11}-\ref{fig13} corresponding respectively to
nonrelativistic and relativistic bosons,
one observes similar isotropisation effects in
the distributions of the emission points and CF's,
as well as similar anisotropy transfer from spatial
to momentum components of the phase space emission points.
The interferometry volume is however conserved only approximately,
it increases by $\sim 30 \%$ from the initial value
of $\sim 176~$ fm$^{3}$ to
the final one of $\sim$ 250~ fm$^{3}$.

\subsubsection{\label{sec722}Nonthermal anisotropic momentum distribution}

\begin{figure}
\includegraphics[width=14cm]{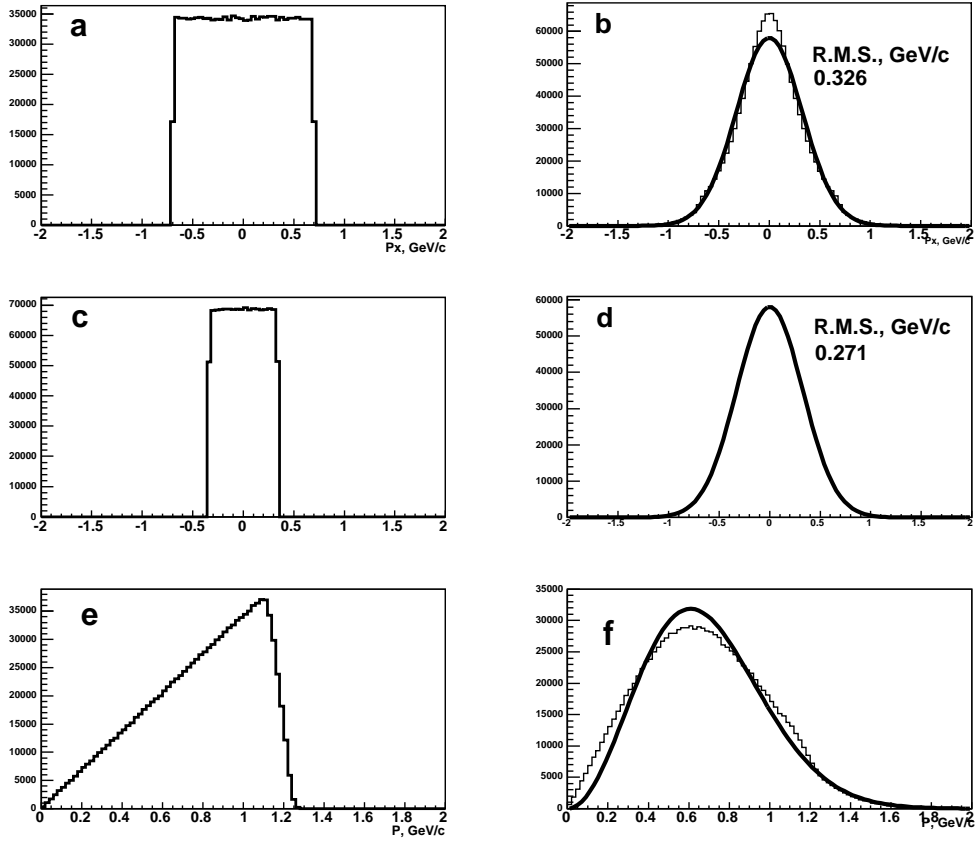}
\caption{\label{fig14}
Final distributions of the components ({\bf b}, {\bf d}) and
the magnitude ({\bf f}) of particle three--momentum obtained
in the UKM-R simulation with $\sigma_{el}=$ 400 mb
on the same conditions as in Fig.~\ref{fig9} except for the initial
three--momentum distribution ({\bf a}, {\bf c}, {\bf e}) defined by
$|p_x| \le 0.7$~GeV/c, $|p_y|,|p_z|  \le 0.35$~GeV/c.
The fitted Gaussian widths of final distributions are shown in
panels {\bf b}, {\bf d} and {\bf f}.
 }
\end{figure}

\begin{figure}
\includegraphics[width=14cm]{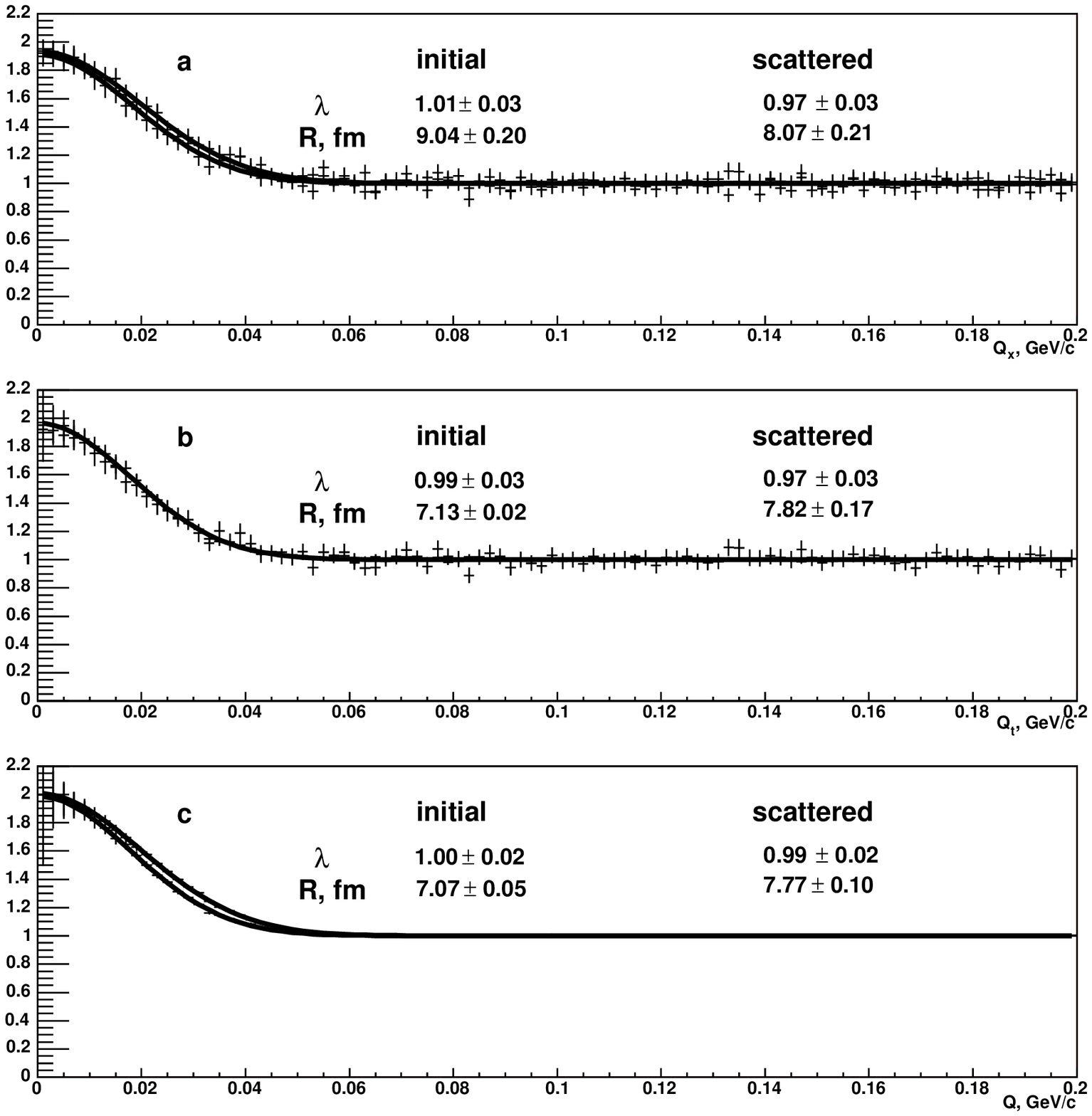}
\caption{\label{fig15}
Initial and final CF's corresponding to the UKM-R
simulation specified in the caption of Fig.~\ref{fig14}.
The CF's are shown as functions of $q_x$ ($q_t \le 10$~MeV/c),
$q_t = ({q_y}^2+{q_z}^2)^{1/2}$ ($q_x \le 7$~MeV/c)
and ${\rm q}=|{\bf q}|$.
 }
\end{figure}

We have further considered the same initial conditions as in
Fig.~\ref{fig9} except for the thermal momentum
distribution which has been substituted by a uniformly populated
parallelepiped defined by
$|p_x| \le 0.7$~GeV/c, $|p_y|, |p_z| \le 0.35$~GeV
(Fig.~\ref{fig14}).
The results of the kinetic evolution of the three-momentum
distribution and the CF's
are shown in figures \ref{fig14} and \ref{fig15}.
One may see that the collisions force the initially
nonthermal distribution to become close to the thermal one.
The final interferometry radii are $\sim 10\%$ higher
than the initial one of 7 fm, similar to the case of the
initial thermal momentum distribution.

\section{\label{sec8} Conclusions and outlook }

We have studied numerical solutions of the Boltzmann
equation in the Universal Kinetic Model.
First, we have tested the model kinetic algorithm recovering the
known nonrelativistic solution [1-3]
for the spherically symmetric Gaussian initial conditions
up to the elastic cross section of 1000~mb.
Particularly, we have checked that despite a huge number
of collisions
the momentum spectra and interferometry radii
remain unchanged during the evolution.
Further, we have modified the initial conditions and found
that with the increasing elastic cross section the system
of nonrelativistic particles more and more recovers
spherical symmetry and thermal momentum distribution.
At the same time, similar to the above case, the momentum
dispersion and the interferometry volume is practically
conserved during the evolution.
The conservation effect takes place also in the evolution
of the system of relativistic particles, except for some
increase
($\sim 30\%$ at 400 mb) of the final interferometry volume.
Our studies thus support similar results
obtained in hydrodynamic approximation in Ref. \cite{akksin}.

It is worth to note that the initially
anisotropic system tends to an isotropic one and, after sufficiently
large number of collisions, the spectra and interferometry radii
are approximately
conserved in subsequent evolution.
In such a case one can exploit
the idea of duality of hydrodynamic and kinetic descriptions of
interferometry volumes and effective temperatures
(momentum dispersions) \cite{Borysova}, i.e.
estimate their values from the thermal hadronic distribution
function in the initially formed hadronic fireball
instead of a detailed analysis based on the complicated
asymptotic distribution or emission functions.

We plan to extend the analysis of kinetic evolution of the
spectra and interferometry radii to many component hadron gases,
including inelastic collisions and resonance decays.

\begin{acknowledgments}
The research has been carried out within the scope of the ERG (GDRE):
Heavy ions at ultrarelativistic energies -
a European Research Group comprising
IN2P3/CNRS, Ecole des Mines de Nantes, Universite de Nantes,
Warsaw University of Technology, JINR Dubna, ITEP Moscow and
Bogolyubov Institute for Theoretical Physics NAS of Ukraine.
It was supported by the
Grant Agency of the Czech Republic under contract 202/04/0793
and by Award No. UKP1-2613-KV-04 of the U.S. Civilian
Research $\&$ Development Foundation for the Independent
States of the Former Soviet Union (CRDF),
Grant of Russian Agency for Science and Innovations 
under contract 02.434.11.7074 (2005-RI-12.0/004/022).
\end{acknowledgments}

\vfill\eject

\clearpage

\end{document}